\documentclass[%
 reprint,
showpacs,
 amsmath,amssymb,
 aps,
]{revtex4-1}

\usepackage{graphicx}
\usepackage{dcolumn}
\usepackage{bm}
\usepackage{bbm}
\usepackage{braket}
\usepackage{float}
\usepackage{chngcntr}
\usepackage[usenames]{color}

\begin{document}


\title{Nonlinear evolution of fast neutrino flavor conversion in the preshock region of core-collapse supernovae}

\author{Masamichi Zaizen$^{1}$}
\email{mzaizen@astron.s.u-tokyo.ac.jp}
\author{Taiki Morinaga$^{2}$}

\affiliation{%
$^{1}$Department of Astronomy, Graduate School of Science, University of Tokyo, Tokyo 113-0033, Japan \\
$^{2}$Graduate School of Advanced Science and Engineering, Waseda University, 3-4-1 Okubo, Shinjuku, Tokyo 169-8555, Japan
}%

\date{\today}

\begin{abstract}
In environments with high dense neutrino gases, such as in core-collapse supernovae, the neutrinos can experience collective neutrino oscillation due to their self-interactions.
In particular, fast flavor conversion driven by the crossings in the neutrino angular distribution can affect explosion mechanism, nucleosynthesis, and neutrino observation.
We perform the numerical computation of nonlinear flavor evolution on the neutrino angular distribution with tiny crossings expected to be generated in the preshock region.
We demonstrate that the fast instability is triggered and a cascade develops under a realistic three-flavor model considering muon production and weak magnetism in the SN dynamics.
The tiny crossing excites specific spatial modes, and then the flavor instability propagates into other modes which otherwise remain stable due to the nonlinear effects.
Our results indicate that fast flavor conversion can rise in the preshock region and have a sufficient impact on the flavor contents.
\end{abstract}

\maketitle


\section{Introduction}
Massive stars experience core collapse and end their lives.
An enormous amount of neutrinos is released from the core-collapse supernovae (CCSNe) and enables us to cultivate the understanding of the inner physics from the observation \cite{Langanke:2003,Mezzacappa:2005,Woosley:2005,Kotake:2006,Janka:2012,Janka:2017,Burrows:2013,Foglizzo:2015,Scholberg:2012,Horiuchi:2018}.
On the other hand, the theoretical understanding of neutrino flavor conversions, especially collective neutrino oscillation, is also essential in the context of core-collapse supernovae \cite{Duan:2010,Mirizzi:2016,Chakraborty:2016}.
Collective neutrino oscillation is a nonlinear phenomenon induced by neutrino-neutrino interactions and can lead to flavor mixing near the core.
If collective flavor conversions occur inside CCSNe, they can significantly influence neutrino reactions during propagating through the medium.
In particular, flavor conversions can enhance a neutrino heating process, in which emitted neutrinos depositing the energy to the stalled shock wave help to succeed the explosion.

Many studies about collective neutrino oscillation have been performed based on the simplified symmetric model, called the bulb model \cite{Duan:2006b}.
The bulb model demonstrates dramatic flavor conversions such as spectral splits \cite{Fogli:2007,Dasgupta:2008b}, while it has also revealed the possibility that the dense matter profile in CCSNe can suppress the collective effects completely \cite{Esteban-Pretel:2008,Banerjee:2011,Chakraborty:2011,Chakraborty:2011a,Zaizen:2018}.
The symmetry breaking in space-time can overcome the matter suppression and lead to the collective flavor conversions \cite{Raffelt:2013,Mirizzi:2013,Chakraborty:2014,Zaizen:2021,Chakraborty:2016b,Dasgupta:2015,Capozzi:2016}.

Recently it has been discussed that a flavor instability, called fast flavor conversion, can occur even under much higher matter density environments \cite{Sawyer:2005,Sawyer:2016,Chakraborty:2016a,Dasgupta:2017,Izaguirre:2017,Capozzi:2017,Dasgupta:2018,Abbar:2018,Airen:2018,Abbar:2019,Nagakura:2019,DelfanAzari:2020,Abbar:2020,Glas:2020,Morinaga:2020,Capozzi:2020,Morinaga:2021a}.
The oscillation modes are fast enough to ignore the vacuum frequency $\omega=\Delta m/2E \sim \mathcal{O}(1)\mathrm{~km^{-1}}$ for a $10\mathrm{~MeV}$ neutrinos, and the flavor instability is induced only by the self-interaction potential.
The fast modes develop on scales $\mu^{-1}\sim (\sqrt{2}G_{\mathrm{F}}n_{\nu})^{-1} \lesssim \mathcal{O}(1)\mathrm{~m}$, depending on the neutrino number density.
The existence of fast flavor conversion is equivalent to the presence of a zero crossing in the neutrino flavor lepton number (NFLN) angular distribution \cite{Morinaga:2021a}.
There has particularly been much discussion of the oscillation possibility using the electron lepton number (ELN) as an indicator.
The region where the ELN crossing can appear has been found near the proto-neutron star, where collective effects are completely suppressed by dense matter in the previous studies based on the bulb model \cite{Abbar:2019,Nagakura:2019,Abbar:2020,DelfanAzari:2020,Glas:2020}.

Morinaga \textit{et al.} \cite{Morinaga:2020} pointed out that the inward-going components due to the coherent neutrino-nucleus scattering create tiny ELN crossings and can lead to the fast flavor conversion irrespective of multidimensional effects in SN dynamics.
Scattered $\bar{\nu}_e$ is enhanced compared to $\nu_e$ due to the difference in the average energy and that it generates the negative crossing in the backward direction.
However, the suggestion is still on the linear stability analysis and the flavor evolution in the nonlinear regime remains to be explored.
Also, Capozzi \textit{et al.} \cite{Capozzi:2020} suggested that the inclusion of $\mu$LN and $\tau$LN provides different results from the effective two-flavor case assuming $\nu_X=\bar{\nu}_X$.
In particular, muon production in SN dynamics can create significant differences in the heavy lepton number and erase the shallow ELN crossings, for example, in the preshock region.

In this paper, we focus on the fast flavor conversion driven by a tiny NFLN crossing in the preshock region.
We perform the nonlinear flavor evolution decomposed into the spatial Fourier modes parallel to the radial direction.
We investigate the impact of the heavy lepton flavor within the three-flavor framework on the occurrence of fast modes.

This paper is organized as follows.
We introduce our numerical strategy decomposing the neutrino kinetic equations into the spatial Fourier modes in Sec.\ref{Sec2},
Also, we present a linear stability analysis and neutrino angular distribution functions to find the dispersion relation of fast modes.
In Sec.\ref{Sec3}, we present the numerical results for fast flavor conversion in effective two-flavor and three-flavor cases.
We summarize our results and the conclusion in Sec.\ref{Sec4}.
\\

\section{Formulation}
\label{Sec2}
\subsection{Evolution of Fourier modes}
The flavor evolution can be described by the equation of motion (EoM) for neutrino density matrices $\rho$ at space-time position $(t,\boldsymbol{x})$,
\begin{equation}
    \mathrm{i}\left(\partial_t +\boldsymbol{v}\cdot\nabla\right)\rho(t,\boldsymbol{x},\Gamma) = \left[H(t,\boldsymbol{x},\Gamma), \rho(t,\boldsymbol{x},\Gamma)\right],
\end{equation}
where $\Gamma$ specifies the neutrino energy $E$ and the flight angle $\boldsymbol{v}$.
The phase-space integration is $\int\mathrm{d}\Gamma^{\prime} = \int^{+\infty}_{-\infty}\mathrm{d}E^{\prime} {E^{\prime}}^2\int\mathrm{d}\boldsymbol{v}^{\prime}/(2\pi)^3$ and also covers over antineutrinos with the flavor-isospin convention~\cite{Airen:2018}, in which antineutrinos are denoted by negative energy as $\overline{\rho}(E)\equiv-\rho(-E)$.
The Hamiltonian $H(t,\boldsymbol{x},\Gamma)$ is
\begin{eqnarray}
    H(t,\boldsymbol{x},\Gamma) &&= U\frac{M^2}{2E}U^{\dagger} + v^{\mu}\Lambda_{\mu} \notag\\
    &&~~~+ \sqrt{2}G_{\mathrm{F}}\int\mathrm{d}\Gamma^{\prime}v^{\mu}v_{\mu}^{\prime}\rho^{\prime},
\end{eqnarray}
which expresses three types of neutrino oscillations: vacuum, matter, and collective neutrino oscillation.
$M^2$ in the first term is the mass-squared matrix and $U$ is the Pontecorvo-Maki-Nakagawa-Sakata matrix.
The second term induces matter oscillation, where $v^{\mu}=(1,\boldsymbol{v})$ and $\Lambda^{\mu} = \sqrt{2}G_{\mathrm{F}}~\mathrm{diag}[\{j^{\mu}_{\alpha}\}]$, with $j^{\mu}_{\alpha}$ being the lepton number current of charged lepton $\alpha$.
The third term corresponds to collective neutrino oscillation induced by the neutrino self-interaction.

We can expand the density matrix and the Hamiltonian by the linear combination of the Pauli matrices $\boldsymbol{\sigma}$ (or the Gell-Mann matrices in the three-flavor case) as
\begin{eqnarray}
    \rho^{n\times n} = \frac{\mathrm{Tr}(\rho)}{n}I_n + \frac{1}{2}\boldsymbol{P}\cdot \boldsymbol{\sigma},
\end{eqnarray}
where $n$ is the number of flavors.
The coefficients of the density matrix are particularly called the polarization vector $\boldsymbol{P}_{\omega,\boldsymbol{v}}$, and the EoM can be recast to
\begin{eqnarray}
    (\partial_t +\boldsymbol{v}\cdot\nabla)\boldsymbol{P}_{\omega,\boldsymbol{v}} &&= (\omega_\mathrm{V}\boldsymbol{B}+\lambda\boldsymbol{L})\times\boldsymbol{P}_{\omega,\boldsymbol{v}} \notag\\
    &&~+\sqrt{2}G_{\mathrm{F}}\int\mathrm{d}\Gamma^{\prime}v^{\mu}v_{\mu}^{\prime}\boldsymbol{P}_{\omega^{\prime},\boldsymbol{v}^{\prime}}\times \boldsymbol{P}_{\omega,\boldsymbol{v}},
\end{eqnarray}
where $\omega_{\mathrm{V}}={\Delta m^2}/{2E}$ is a vacuum frequency and $\lambda=\sqrt{2}G_{\mathrm{F}}n_e$ is a matter potential.
Here we ignore positrons and heavy charged leptons, such as muons and tauons.
The vectors $\boldsymbol{B}$ and $\boldsymbol{L}$ are coefficients corresponding to the vacuum term and the matter term, respectively.
The cross product $\times$ is defined by the structure constants $f_{abc}$, e.g.
\begin{eqnarray}
    (\boldsymbol{B}\times\boldsymbol{P})_c = \sum_{a,b}f_{abc}B_a P_b.
\end{eqnarray}

A partial differential equation can be transformed to a tower of ordinary differential equations decomposed by the Fourier expansion.
The polarization vectors can be converted to the spatial Fourier modes $K$ as
\begin{eqnarray}
    \boldsymbol{P}_{\omega,v}(t,x) = \sum_{K}\mathrm{e}^{\mathrm{i}Kx}\tilde{\boldsymbol{P}}^{K}_{\omega,v}(t).
\end{eqnarray}
Here we set spatial dimension only in the radial direction, and consider spatial modes $\boldsymbol{K}$ parallel to the radial direction, as in Ref.~\cite{Morinaga:2020}.
Also, we assume the azimuthal symmetry in the flight direction.
Still in a nonlinear regime, the EoM is expressed by
\begin{eqnarray}
    \frac{\mathrm{d}}{\mathrm{d}t}\tilde{\boldsymbol{P}}_{\omega,v}^{K} &&= -\mathrm{i}vK\tilde{\boldsymbol{P}}_{\omega,v}^{K} + (\omega_{\mathrm{V}}\boldsymbol{B}+\lambda\boldsymbol{L})\times\tilde{\boldsymbol{P}}_{\omega,v}^{K} \notag\\
    &&~+ \sum_{K^{\prime}}\left[\sqrt{2}G_{\mathrm{F}}\int\mathrm{d}\Gamma^{\prime}v^{\mu}v_{\mu}^{\prime}\tilde{\boldsymbol{P}}_{\omega^{\prime},v^{\prime}}^{K-K^{\prime}}\times\tilde{\boldsymbol{P}}_{\omega,v}^{K^{\prime}}\right].
\end{eqnarray}
The tower has a convolution term for the spatial modes $K$, and it can induce a cascade in the Fourier space.
Here we discretize the spatial modes as $K=n_K K_0 = n_K\times 10~\omega_{\mathrm{V}}$ to make them dimensionless and take them up to $n_K=300$.
We keep $n_K>200$ empty to avoid a spurious rise \cite{Boyd:2001}, following the dealiasing approach in Refs.~\cite{Dasgupta:2015, Capozzi:2016}.
We also take initial perturbations of $\mathcal{O}(10^{-12})$ to seed the spatial inhomogeneity.
In this study, we take neutrino oscillation parameters as follows.
We consider monochromatic energy spectra with the vacuum frequency $\omega_{\mathrm{V}}=6.6\times 10^{-4}\mathrm{~m^{-1}}$ and the mass-squared difference ratio $\epsilon=\Delta m_{21}^2/\Delta m_{31}^2\sim 0.03$ within the three-flavor framework, and we assume normal mass ordering.
Also, we set mixing angles $\theta_{12}=\theta_{13}=\theta_{23}=10^{-3}$ to mimic the suppression due to matter oscillation, and we set $\lambda = 0$.
\\

\subsection{Linear stability analysis}
The linear stability analysis has been used to investigate whether fast flavor conversion can occur.
We will briefly summarize the procedure.
We consider the small perturbation in the off-diagonal components $\rho^{\alpha\beta}$ from the flavor eigenstate $\rho = \operatorname{diag}(\{f_{\nu_\alpha}\})$, with the occupation number of $\alpha$ neutrinos $f_{\nu_\alpha}$.
By neglecting $M^2$, the EoM can be recast to
\begin{equation}
    \mathrm{i}(\partial_t+\boldsymbol{v}\cdot\nabla)\rho_{\boldsymbol{v}}^{\alpha\beta} = v^\mu(\Lambda_{\mu}^{\alpha\beta}+\Phi_{\mu}^{\alpha\beta})\rho_{\boldsymbol{v}}^{\alpha\beta} - \int\frac{\mathrm{d}\boldsymbol{v}^{\prime}}{4\pi}v^{\mu}v_{\mu}^{\prime}G_{\boldsymbol{v}^{\prime}}^{\alpha\beta}\rho_{\boldsymbol{v}^{\prime}}^{\alpha\beta}
\end{equation}
up to the linear order of $\rho^{\alpha\beta}$, where
\begin{equation}
    G^{\alpha}_{\boldsymbol{v}} = \sqrt{2}G_{\mathrm{F}}\int \frac{E^2\mathrm{d}E}{2\pi^2}f_{\nu_\alpha}(\Gamma)
\end{equation}
is the $\alpha$LN angular distribution, $G_{\boldsymbol{v}}^{\alpha\beta} \equiv G_{\boldsymbol{v}}^{\alpha}-G_{\boldsymbol{v}}^{\beta}$, $\Phi_\mu^{\alpha\beta} \equiv \int d^2\boldsymbol{v}/(4\pi) G^{\alpha\beta}_{\boldsymbol{v}}v_\mu$ and $\Lambda_\mu^{\alpha\beta} \equiv \Lambda_\mu^{\alpha\alpha} - \Lambda_\mu^{\beta\beta}$.
We can derive the dispersion relation by substituting the plane wave ansatz $\rho_{\boldsymbol{v}}^{\alpha\beta}=Q_{\boldsymbol{v}}^{\alpha\beta}~\mathrm{e}^{-\mathrm{i}(\Omega t-\boldsymbol{K}\cdot\boldsymbol{x})}$ as
\begin{equation}
    \mathrm{det}\left[\Pi^{\alpha\beta}(k)\right] = 0,
    \label{eq:DR}
\end{equation}
where $k=K-(\Lambda+\Phi)$ and
\begin{equation}
    \Pi^{\alpha\beta\mu\nu}(k) = \eta^{\mu\nu} + \int\frac{\mathrm{d}\boldsymbol{v}}{4\pi}G_{\boldsymbol{v}}^{\alpha\beta}\frac{v^{\mu}v^{\nu}}{v\cdot k}.
\end{equation}
Therefore, the dispersion relation is determined by the difference between the NFLN angular distributions of two flavors $G^{\alpha\beta}_{\boldsymbol{v}}$.
When $f_{\nu_X} = f_{\bar{\nu}_X}$ for heavy lepton-flavor neutrinos $\nu_X$, the dispersion relation is nontrivial only for the $e$-$X$ sector and depends only on the ELN angular distribution $G^e_{\boldsymbol{v}}$.

If there is a nonreal $\Omega$ for some real vector $\boldsymbol{K}$, it means instability and perturbations will grow exponentially.
In the above discussion, we omit the vacuum mixing term.
The instability under this assumption is called fast instability, while the instability appearing when this assumption is relaxed is called slow instability~\cite{Airen:2018}.
The slow instability may be important when we focus on the instability in the preshock region, in which the vacuum mixing term is comparable to the self-interaction term of neutrinos, as we will show.
\\

\subsection{Angular distribution}
\begin{figure}[b]
    \centering
    \includegraphics[width=0.9\linewidth]{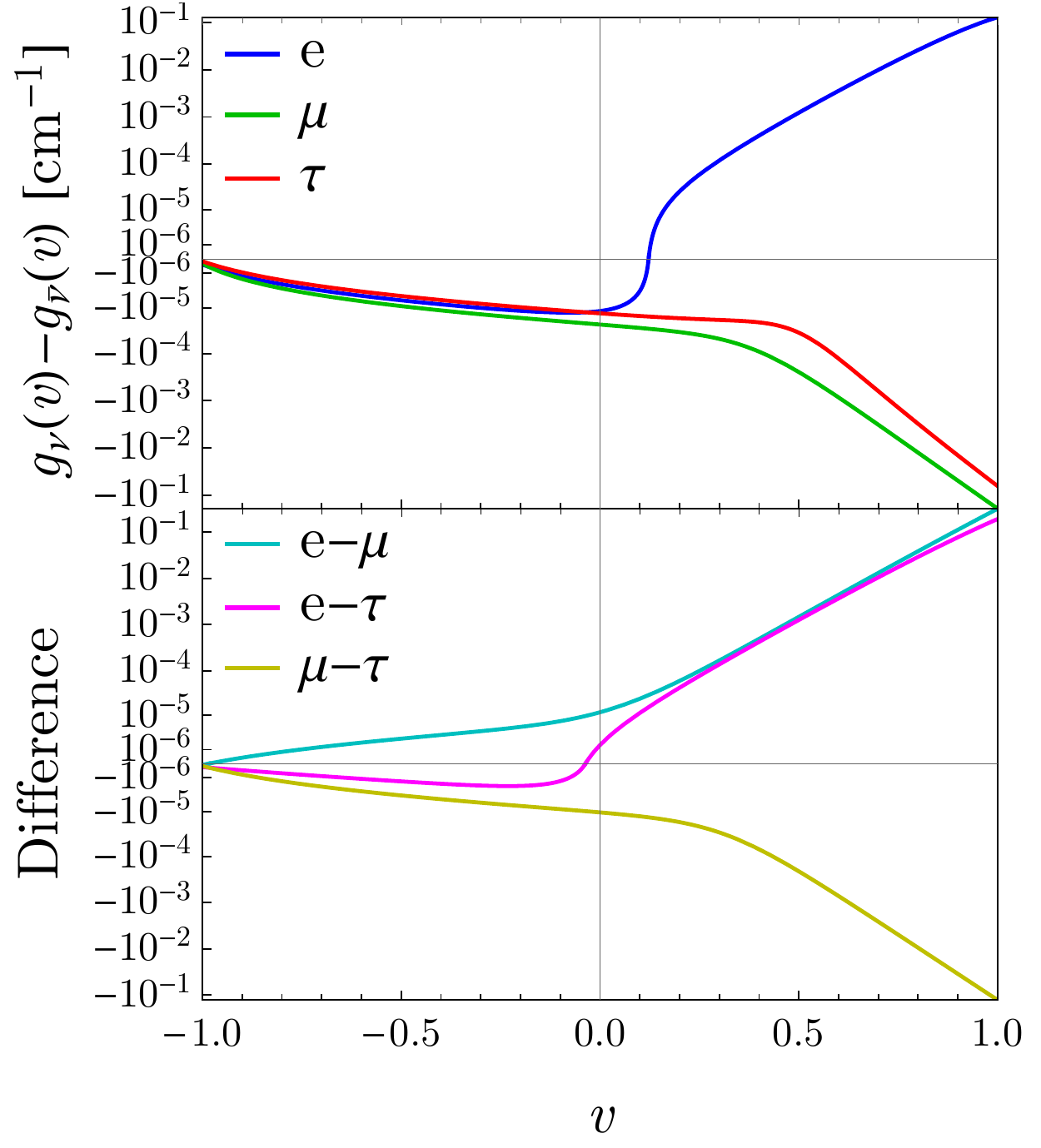}
    \caption{Top: The NFLN angular distribution $G_{\boldsymbol{v}}^{\alpha}$ for each flavor $\alpha$.
    The ELN angular distribution has a zero crossing, while the $\mu$LN and $\tau$LN angular distributions are always negative due to the effects of muon production and weak magnetism.
    Bottom: The difference $G_{\boldsymbol{v}}^{\alpha\beta}$ between the NFLN angular distribution for two flavors $\alpha$ and $\beta$.
    The emission of muon antineutrinos enhanced by muon production erases the ELN crossing, while a crossing still survives only in the $e-\tau$ sector.
    In both panels, the vertical axes are scaled by the function $f(x) = \operatorname{sgn}(x)\log(1 + 10^6 x)$.
    }
    \label{fig:NFLN}
\end{figure}

The coherent backward scattering off heavy nuclei can create a zero crossing of the angular distribution in the preshock region.
\begin{table}[b]
    \centering
    \begin{tabular}{ccccc}\hline\hline
          Flavor & $g_b~(10^{26}\mathrm{cm^{-3}})$ & $g_b^{\prime}~(10^{27}\mathrm{cm^{-3}})$& $g_f~(10^{32}\mathrm{cm^{-3}})$ & $b~(10^{5})$ \\ \hline
          $\nu_{e}$ & $4$ & $4.5$ & $2.7$ & $1.5$ \\
          $\bar{\nu}_{e}$ & $4.5$ & $6$ & $2.5$ & $2.5$ \\
          $\nu_{\mu/\tau}$ & $3$ & $6$ & $1.5$ & $4$ \\
          $\bar{\nu}_{\mu}$ & $3.4$ & $8.1$ & $1.8$ & $4.5$ \\
          $\bar{\nu}_{\tau}$ & $3.2$ & $7.2$ & $1.6$ & $4.5$ \\
          \hline\hline
    \end{tabular}
    \caption{Parameters in the radial angular distribution function in Eq.(\ref{eq:dist}).}
    \label{tab:dist}
\end{table}
To investigate the nonlinear evolutions induced by these crossings, we employ the following model for the initial condition of the radial angular distribution $g(v)$, which is the average of $G^\alpha_{\boldsymbol{v}}$ over the azimuthal angle and $v=\cos\theta$:
\begin{equation}
    g(v) = g_b + g_b^{\prime}(\mathrm{e}^{v+1}-1) + g_f b^{v-1},
    \label{eq:dist}
\end{equation}
where the parameters $g_b, g_b^{\prime}, g_f$, and $b$ for each flavor are chosen as shown in Table~\ref{tab:dist} ($g(v)$ here treats neutrinos and antineutrinos independently) and the resultant NFLN angular distributions and their differences are shown in Fig.~\ref{fig:NFLN}.
This model is designed to reproduce the strongly forward-peaked distribution in the preshock region.
As a matter of fact, when $g_f \gg g_b, g_b'$ and $b^{-2} \ll g_b/g_f, g_b'/g_f$ are satisfied, the asymptotic behaviors of $g(v)$ are
\begin{eqnarray}
    g(v) \sim 
    \begin{cases}
    g_f b^{v-1} &\mathrm{~for~}v\sim 1 \\
    g_b +g_b^{\prime}(v+1) &\mathrm{~for~}v\sim -1
    \end{cases}
\end{eqnarray}
and $g_f$ is the intensity at $v = 1$;
$b$ determines the sharpness of the forward peak and is considered to be larger for the smaller radius of neutrinosphere $R_\nu$ (subscript $\nu$ is sometimes replaced by $\nu_e$, for example, to denote $R_\nu$ for $\nu_e$ henceforth);
$g_b$ is the intensity at $v = -1$ and is proportional to $L_\nu E_\nu R_\nu^{-2}$ in the bulb model in Ref.~\cite{Morinaga:2020}, where $L_\nu$ and $E_\nu$ are luminosity and mean energy, respectively;
$g_b'$ corresponds to the gradient at $v = -1$ and $\propto L_\nu E_\nu$ in the bulb model.
We consider the following physical processes to choose the parameters in Table~\ref{tab:dist}.

First, since $\bar\nu_e$ is decoupled from the matter at a smaller radius than $\nu_e$, $R_{\bar\nu_e} < R_{\nu_e}$ and $E_{\bar\nu_e} > E_{\nu_e}$ are satisfied.
As a result, $g_b$, $g_b'$ and $b$ for $\bar\nu_e$ is larger than $\nu_e$ while $g_f$ for $\bar\nu_e$ is usually smaller than $\nu_e$.
This process causes the ELN crossing as shown in the top panel of Fig.~\ref{fig:NFLN}.
The parameters we choose can indeed almost reproduce the angular distributions for $\nu_e$ and $\bar\nu_e$ in the realistic supernova model in Ref.~\cite{Morinaga:2020}.
Also, the luminosity of the heavy-leptonic neutrinos $\nu_X$ are smaller than $\nu_e$ and $\bar\nu_e$ while they have smaller (larger) $R_\nu$ ($E_\nu$) than $\bar\nu_e$.
As a result, $g_f$ (and possibly $g_b$) for $\nu_X$ is smaller than $\nu_e/\bar\nu_e$, while $g_b'$ and $b$ for $\nu_X$ may be larger.

Based on these ideas, we also take muon production~\cite{Bollig:2017,Fischer:2020} and weak magnetism~\cite{Horowitz:2002} into account.
If the muon is created deep inside the supernova core, $\bar\nu_\mu$ emission is enhanced and $\mu$LN angular distribution is lowered.
This effect may cancel out the negative part of the ELN angular distribution and hence NFLN crossings in the $e-\mu$ sector may be absent.
The cross section of neutral-current scattering with nucleons for neutrinos is larger than that for antineutrinos due to the weak-magnetism correction.
It makes $R_\nu$ for neutrinos larger than antineutrinos and hence $E_\nu$ for neutrinos become smaller than antineutrinos.
The model parameters for antineutrinos reflect more forward focusing and more scattered behaviors than those for neutrinos.
Smaller $R_{\nu}$ produces more forward focusing angular distribution, and larger $E_{\nu}$ reflects the enhancement of the scattering.
Although the effects of muon production and weak magnetism have not been sufficiently clarified to determine the distributions quantitatively, we choose possible parameters that can generate NFLN crossing with reference to Ref.~\cite{Bollig:2017}.
As confirmed in the bottom panel of Fig.~\ref{fig:NFLN}, only $G^{e\tau}_{\boldsymbol{v}}$ has the crossing in our model.
We note that the crossing may disappear if the correction for the heavy lepton-flavor neutrinos is more significant than the model we employ.
One needs to consider more realistic SN simulations to get more accurate neutrino angular distributions.

\begin{figure}[t]
    \centering
    \includegraphics[width=0.9\linewidth]{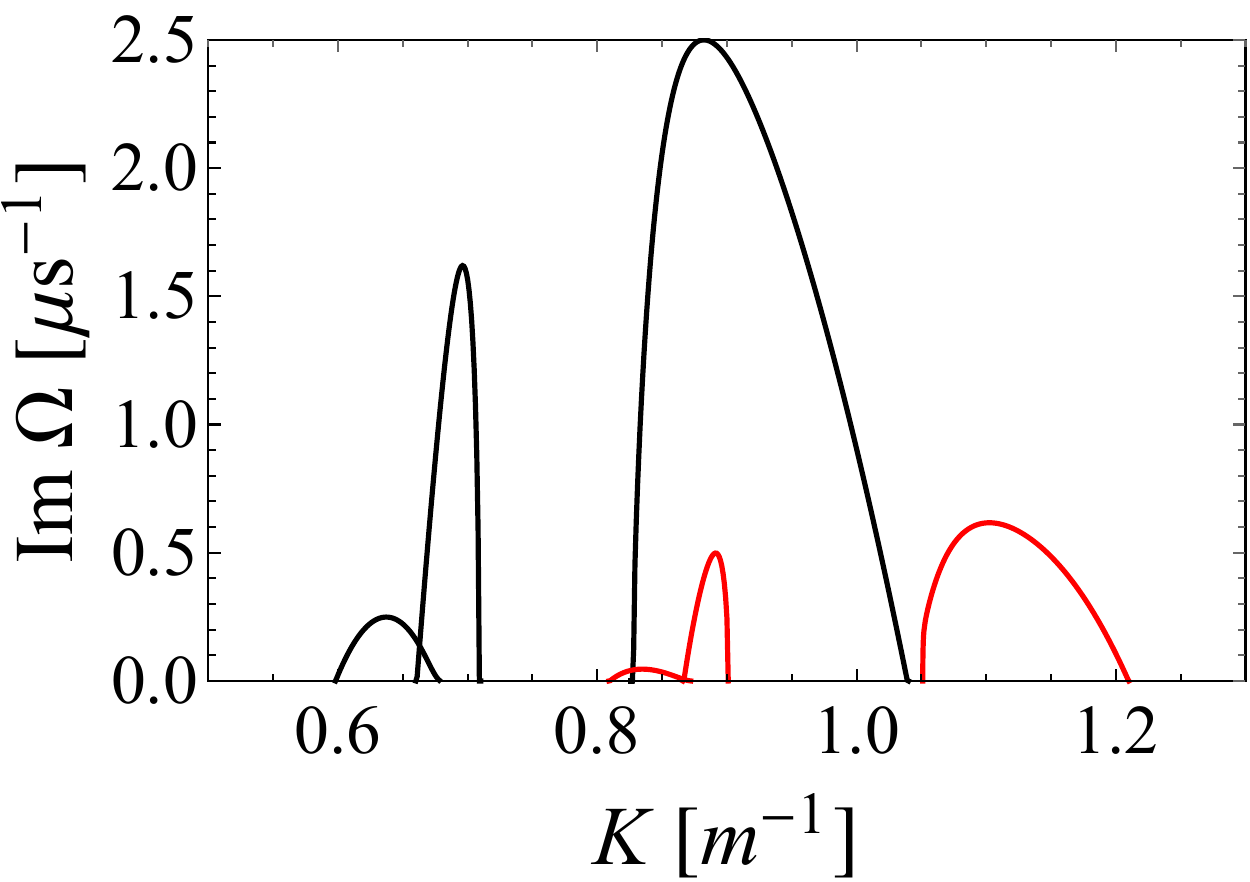}
    \caption{Growth rates $\operatorname{Im}\Omega$ as a function of real $K$.
    The black lines are only for the ELN crossing $G_{\boldsymbol{v}}^{e}$, and the red lines are for the NFLN difference $G_{\boldsymbol{v}}^{e\tau}$ within the three-flavor framework.
    }
    \label{fig:DR}
\end{figure}
Figure~\ref{fig:DR} shows the growth rate $\operatorname{Im}\Omega$ as a function of real $K$.
Black lines are unstable modes only for the ELN crossing within the effective two-flavor framework, and red lines are for the NFLN difference $G_{\boldsymbol{v}}^{e\tau}$ within the three-flavor framework.
There are three branches in the dispersion relation in both cases, and each spatial mode is expected to be excited by the flavor instability in numerical simulation.
The growth rate in the $e-\tau$ sector becomes about four times smaller than that for the ELN crossing within the two-flavor framework because the $\tau$LN angular distribution $G_{\boldsymbol{v}}^{\tau}$ partly weakens the ELN crossing.
Note that we present the dispersion relation after transforming$(\omega,k)$ in Eq.(\ref{eq:DR}) into $(\Omega,K)$, because spatial modes given in numerical simulation are original $K$, not $k$.
\\

\section{Numerical results}
\label{Sec3}
In this section, we present the results of the calculations in the following two cases.
\begin{enumerate}
    \item 
    First, we consider only the ELN crossing within the effective two-flavor case, assuming $f_{\nu_X} = f_{\bar{\nu}_X}$. 
    We drop the vacuum term to confirm the consistency with the linear stability analysis and perform the numerical simulation.
    The vacuum term has a role in generating flavor perturbation, and we take the perturbation of $\mathcal{O}(10^{-12})$ to mimic the role.
    To investigate the effect of the vacuum term, which is disregarded in the fast regime, we also perform a similar calculation including the vacuum term.
    
    \item
    Second, we perform the numerical simulation within the three-flavor framework considering the angular distribution of $\mu$LN and $\tau$LN and ignoring the vacuum term to confirm the consistency.
    Similarly, we also perform the simulation including the vacuum term.
\end{enumerate}

\subsection{Effective two-flavor case only with ELN crossing}
\label{Sec.3A}
\begin{figure}[t]
    \centering
    \includegraphics[width=0.9\linewidth]{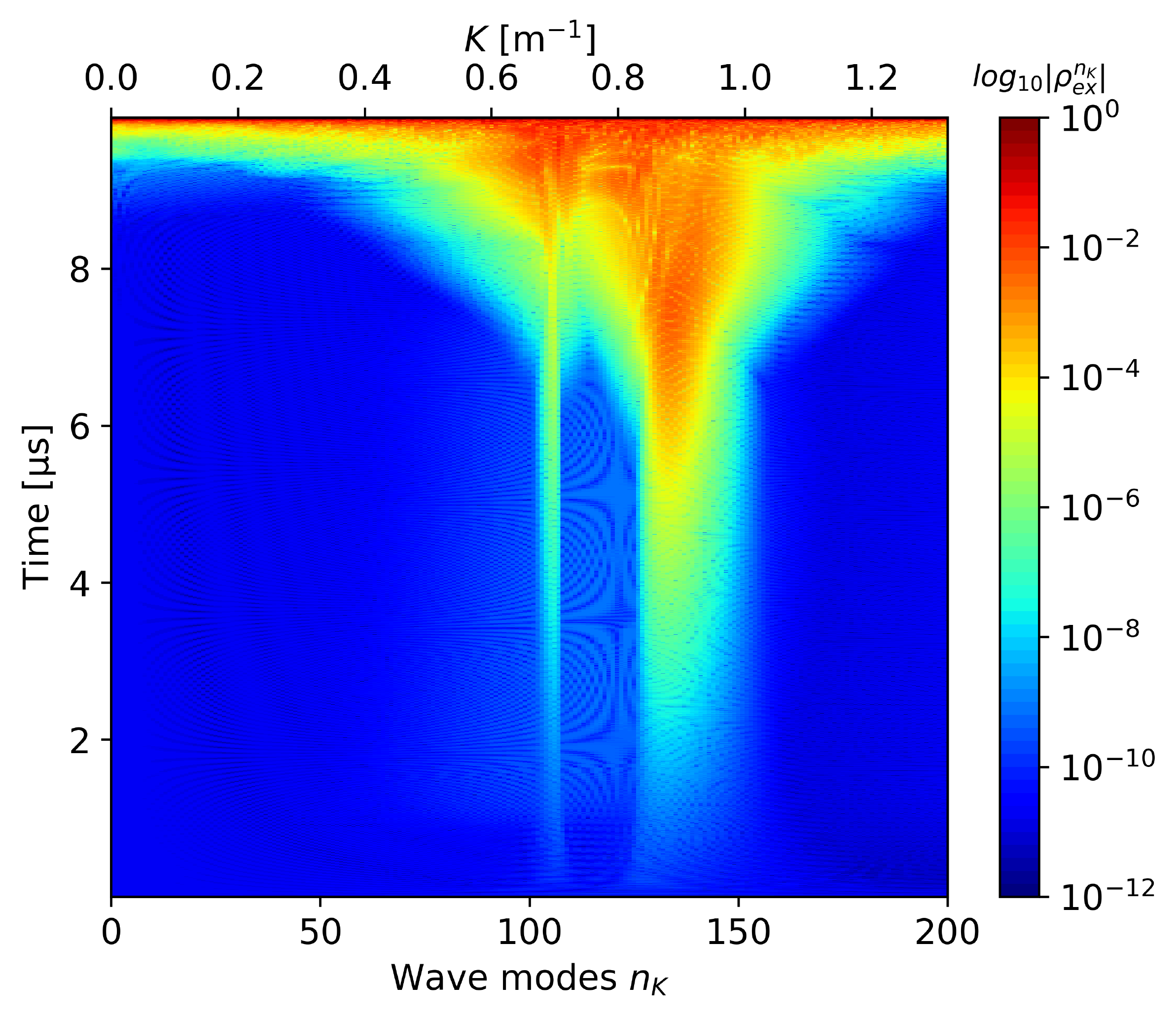}
    \caption{The case neglecting the vacuum term within the effective two-flavor framework.
    The time evolution of the angle-averaged off-diagonal term $|\rho_{ex}^{n_K}|$ for various spatial Fourier modes $K$.
    Spatial modes around $K=0.7\mathrm{~m^{-1}}$ and $0.9\mathrm{~m^{-1}}$ are first excited and then the flavor instabilities spread to different modes due to the nonlinear term after $t\sim 6\mathrm{\mu s}$.
    }
    \label{fig:novac_nK}
\end{figure}

We first present the case omitting the vacuum term in the effective two-flavor case.
Figure~\ref{fig:novac_nK} shows the time evolution of the angle-averaged off-diagonal term $\displaystyle |\rho_{ex}^{n_K}|=\Big\langle\left|(\tilde{\boldsymbol{P}}^{K}_{\omega,v})^{(1)}-\mathrm{i}(\tilde{\boldsymbol{P}}^{K}_{\omega,v})^{(2)}\right|\Big\rangle$.
\begin{figure}[t]
    \centering
    \includegraphics[width=0.9\linewidth]{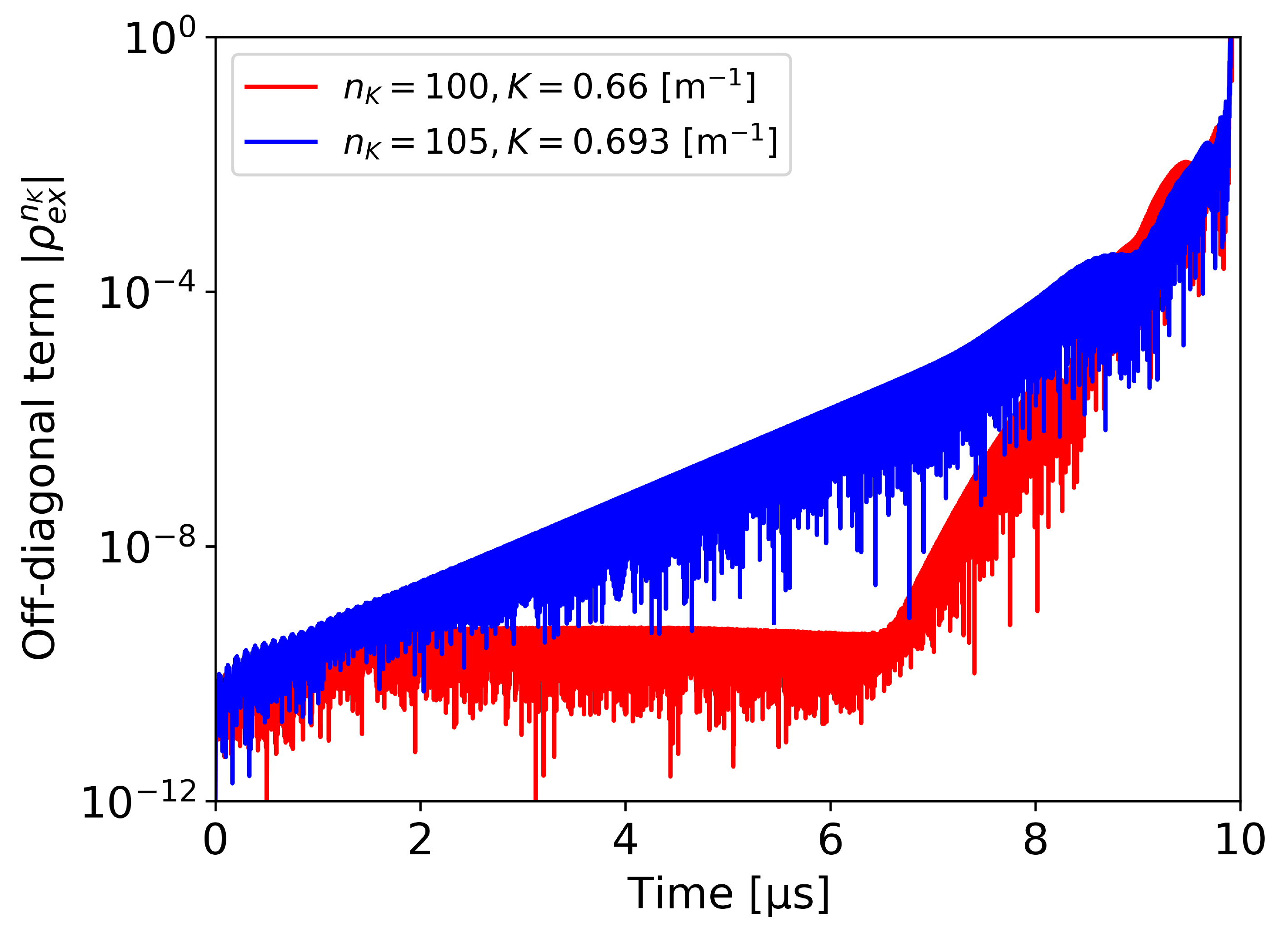}
    \includegraphics[width=0.9\linewidth]{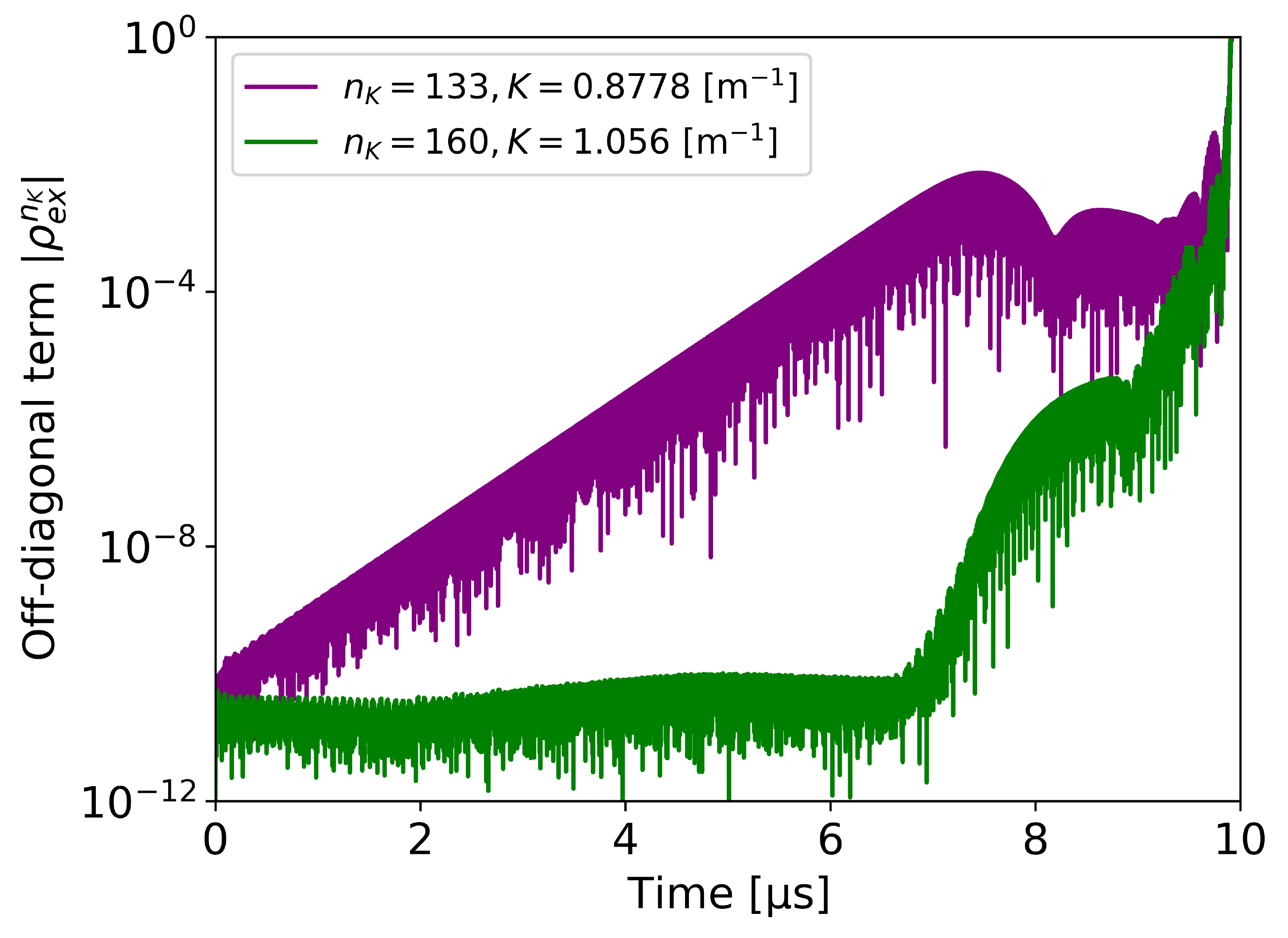}
    \caption{The time evolution of the off-diagonal term $|\rho_{ex}^{n_K}|$ for some spatial Fourier modes in the case omitting the vacuum term.
    The top panel is with $K=0.66\mathrm{~m^{-1}}$ and $0.693\mathrm{~m^{-1}}$, inside and outside the middle branch, respectively.
    The bottom panel is with $K=0.8778\mathrm{~m^{-1}}$ and $1.056\mathrm{~m^{-1}}$, inside and outside the rightmost branch, respectively.
    }
    \label{fig:novac_growth}
\end{figure}
The flavor evolution in Fourier space displays two branches at early time $t<6\mathrm{~\mu s}$.
The spatial Fourier space that the excited modes span is consistent with the two right unstable branches in Fig.~\ref{fig:DR}.
On the other hand, the dispersion relation gives three branches, and the leftmost flavor instability with the smallest growth rate is missing.
The growth rate of the leftmost flavor instability is about $\operatorname{Im}\Omega\sim (4\mathrm{~\mu s})^{-1}$ and an order of magnitude smaller than the peak in the rightmost branch with the largest growth rate.
It means that the initial perturbation grows only about 4.5 times by the critical time $t\sim6\mathrm{~\mu s}$ when the nonlinear effects appear.
As shown in Fig.~\ref{fig:novac_nK}, spatial modes that would otherwise remain stable grow fast via the nonlinear convolution term after $t\sim6\mathrm{~\mu s}$ and a cascade develops into all Fourier modes.
Therefore, the other unstable branches reach the nonlinear regime first and cover up this tiny flavor instability before it grows sufficiently.
Finally, all spatial Fourier modes receive the flavor instability, and complicated oscillation behaviors appear.

Figure~\ref{fig:novac_growth} shows the time evolution of the off-diagonal term $|\rho_{ex}^{n_K}|$ for some spatial modes.
The top panel is for two spatial modes $0.66\mathrm{~m^{-1}}$ and $0.693\mathrm{~m^{-1}}$, and the bottom is for spatial modes $0.8778\mathrm{~m^{-1}}$ and $1.056\mathrm{~m^{-1}}$.
For each panel, one corresponds to an unstable mode (as we can confirm in Fig.~\ref{fig:DR}) and the other to a stable mode.
The growth rates during the linear phase, estimated from Fig.~\ref{fig:novac_growth}, are $\operatorname{Im}\Omega=1.57\mathrm{~\mu s^{-1}}$ and $2.50\mathrm{~\mu s^{-1}}$ for $K=0.693\mathrm{~m^{-1}}$ and $0.8778\mathrm{~m^{-1}}$, respectively.
These growth rates are consistent with the peak of the two branches predicted by the dispersion relation in Fig.~\ref{fig:DR}.
On the other hand, the stable modes in Fig.~\ref{fig:DR} indeed do not grow in the linear phase before $t\sim 6\mathrm{~\mu s}$, while they quickly grow after that due to the nonlinear effect.

\begin{figure}[t]
    \centering
    \includegraphics[width=0.9\linewidth]{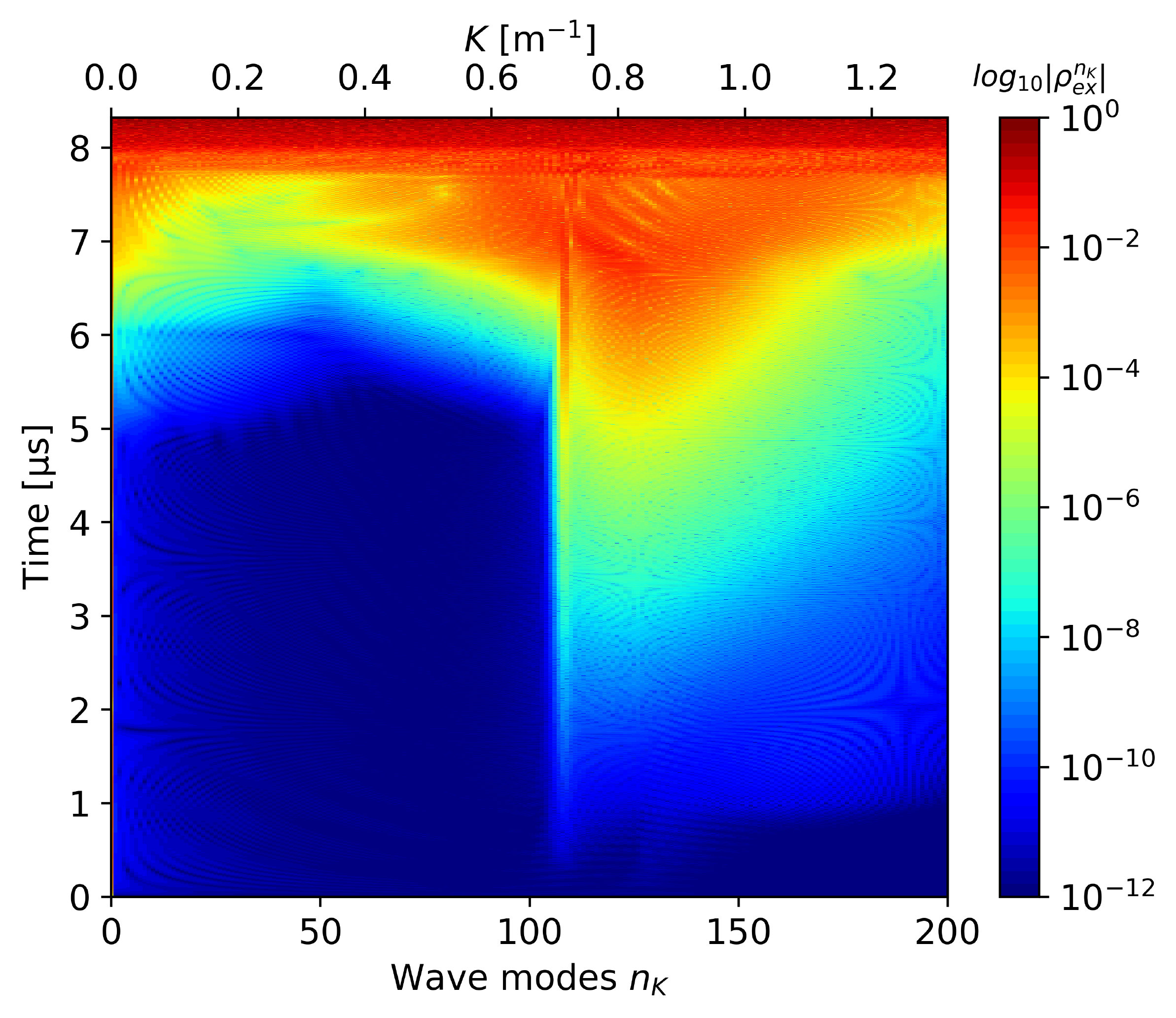}
    \caption{The same as Fig.~\ref{fig:novac_nK}, but for the inclusion of the vacuum term.
    The flavor evolution is largely different from the case omitting the vacuum term and may result from the presence of slow instability.
    }
    \label{fig:2f_nK}
\end{figure}
As a comparison, we perform a similar simulation including the vacuum term and present it in Fig.~\ref{fig:2f_nK}.
In this case, perturbation seeds are naturally given by mixing angles and then lead to fast flavor conversion by the self-interaction potential.
The narrow components corresponding to the middle branch in Fig.~\ref{fig:DR} is prominent, while the modes that seem to correspond to the rightmost branch are much broader, and there is no gap between the two branches.
In the preshock region, the self-interaction potential $\Phi$ is not large enough to neglect the vacuum frequency completely, and the slow instabilities associated with it may influence the evolution of fast modes.
Reference~\cite{Airen:2018} has suggested that unstable modes emerge not only near the origin $k=0$ but also at larger $k$ due to mixing between fast and slow modes in the form of the nonvanishing vacuum term but in the simple colliding beam model.
In our angular distribution model, the spatial mode $K$ of the middle branch in Fig.~\ref{fig:DR} is near the origin, and the mixing between fast and slow modes may rise to fill the gap.
\\

\subsection{ELN+$\mu$LN+$\tau$LN angular distribution}
\begin{figure*}[ht]
    \centering
    \includegraphics[width=0.32\linewidth]{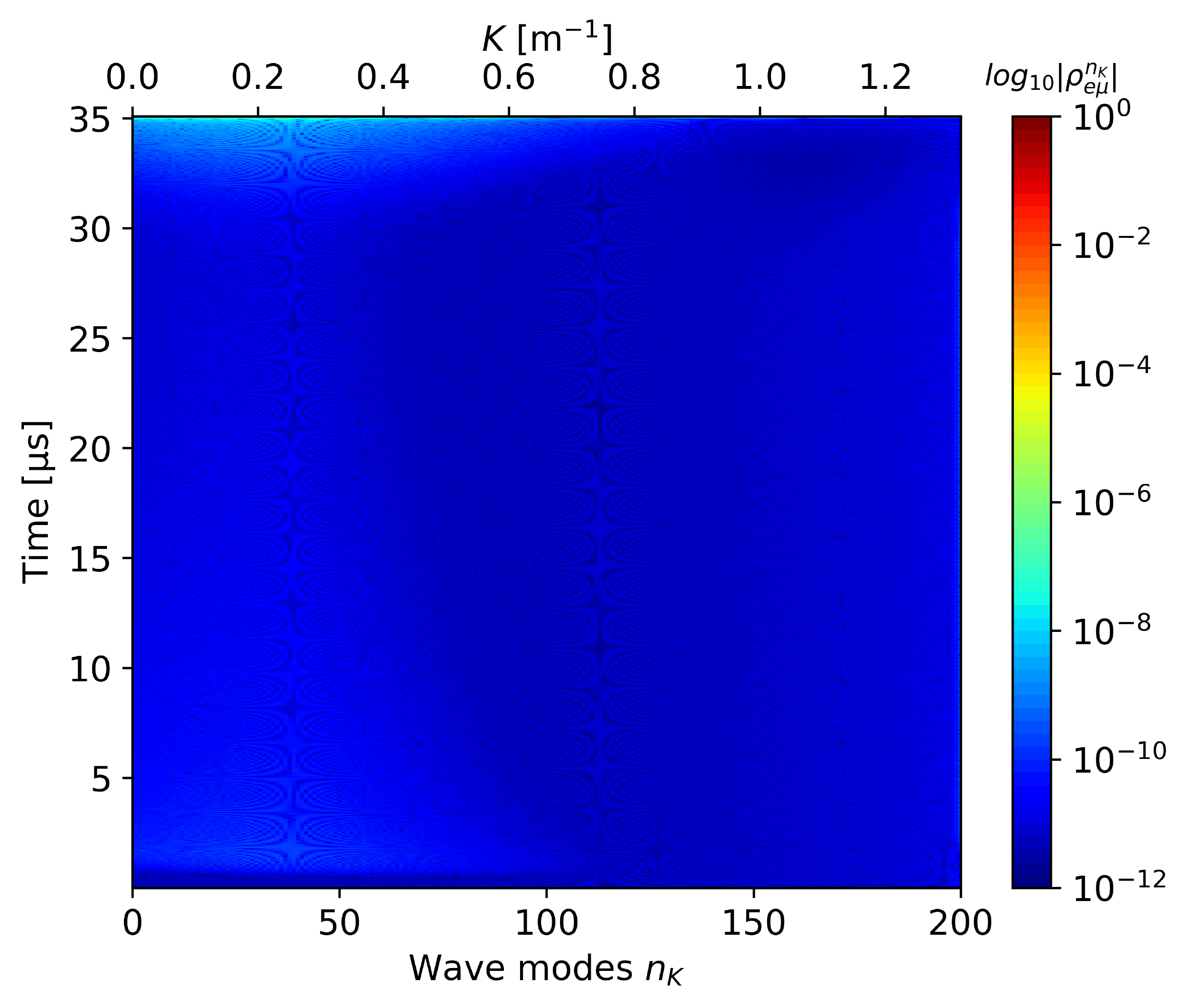}
    \includegraphics[width=0.32\linewidth]{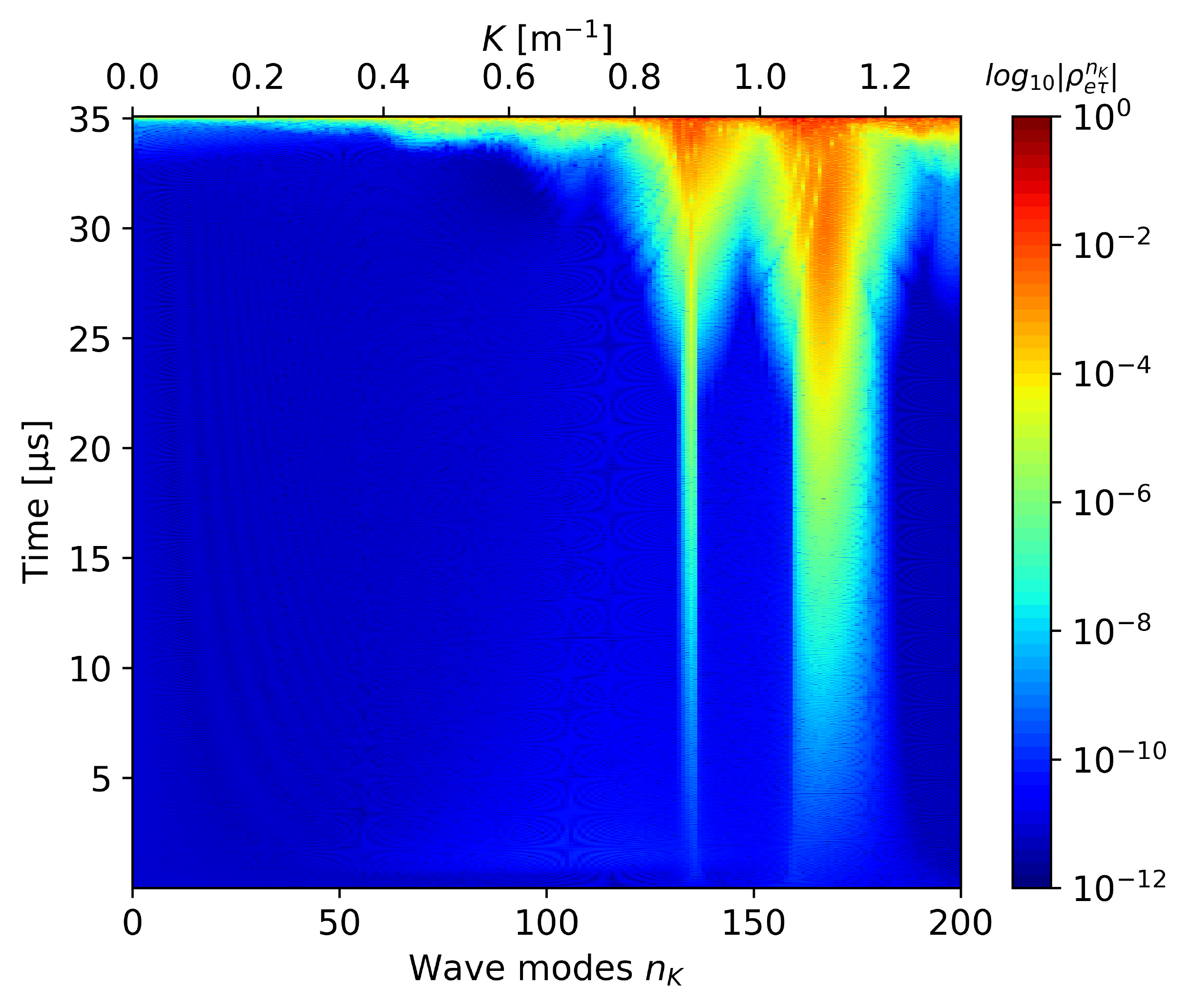}
    \includegraphics[width=0.32\linewidth]{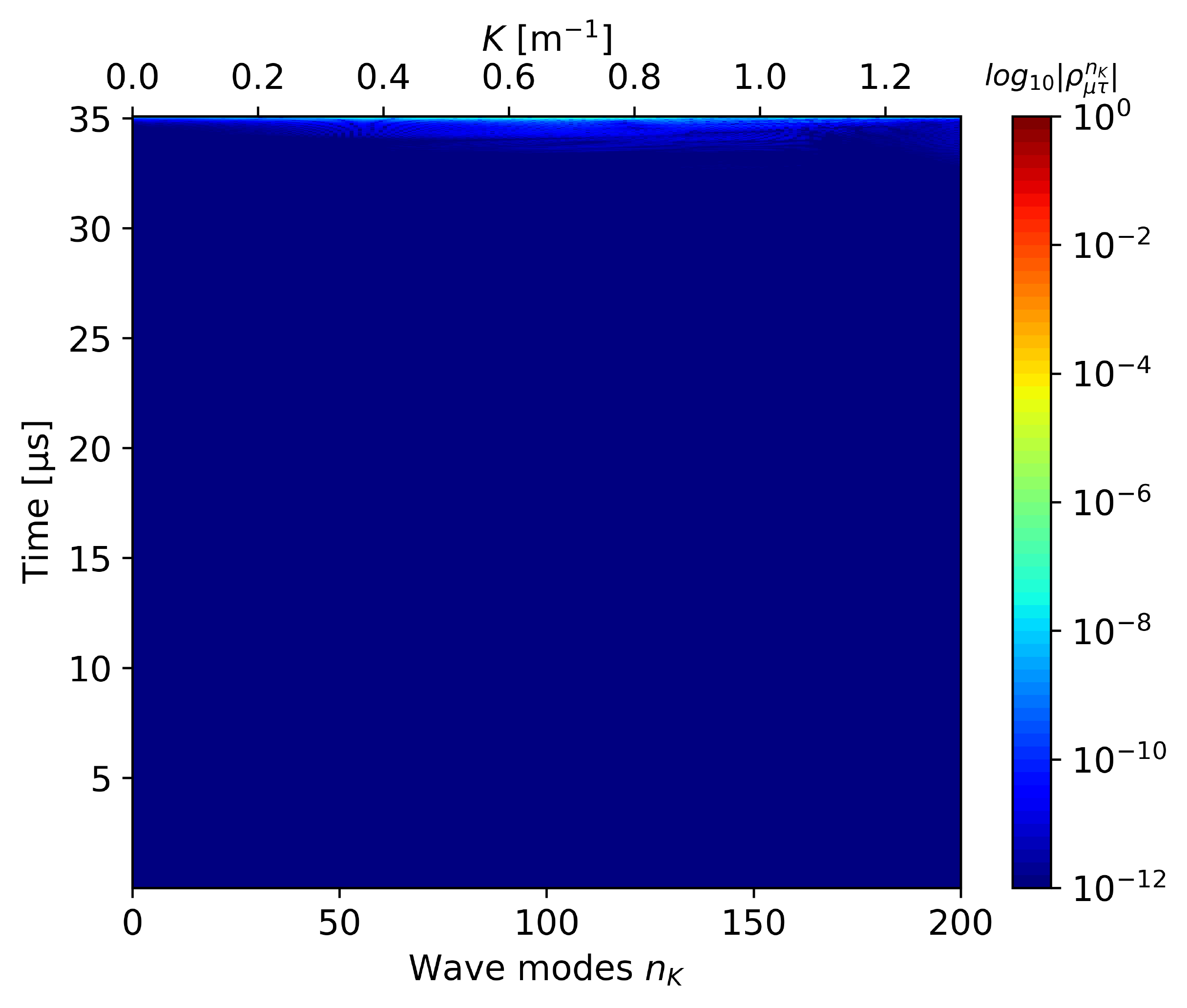}
    \caption{The same as Fig.~\ref{fig:novac_nK}, but for the three-flavor framework.
    From left to right panels: the $e-\mu$, $e-\tau$, and $\mu-\tau$ sectors.
    The flavor instability appears to grow only in the $e-\tau$ sector due to the crossing of the NFLN difference $G_{\boldsymbol{v}}^{e\tau}$.
    }
    \label{fig:3f_novac_nK}
\end{figure*}
\begin{figure*}[ht]
    \centering
    \includegraphics[width=0.32\linewidth]{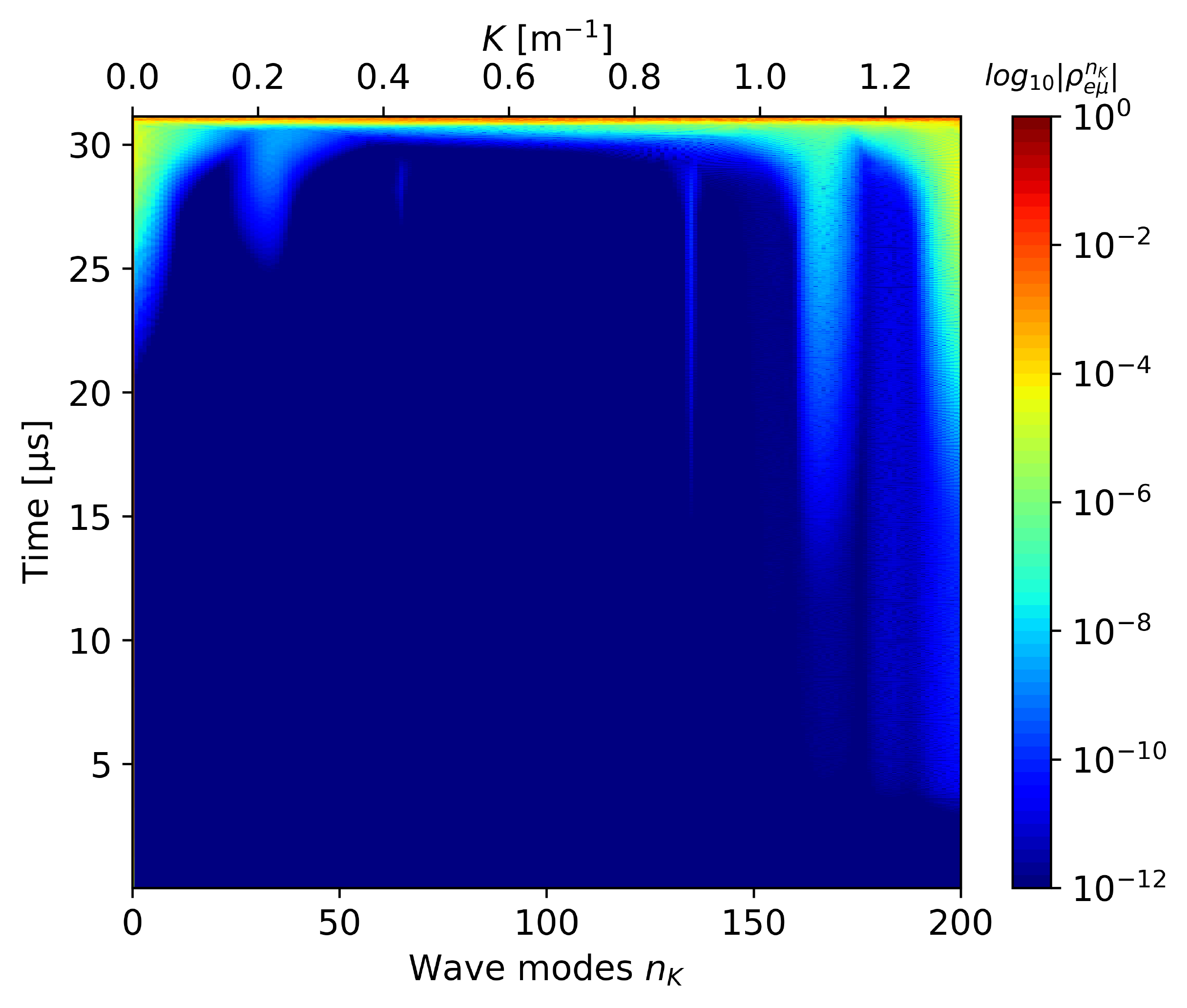}
    \includegraphics[width=0.32\linewidth]{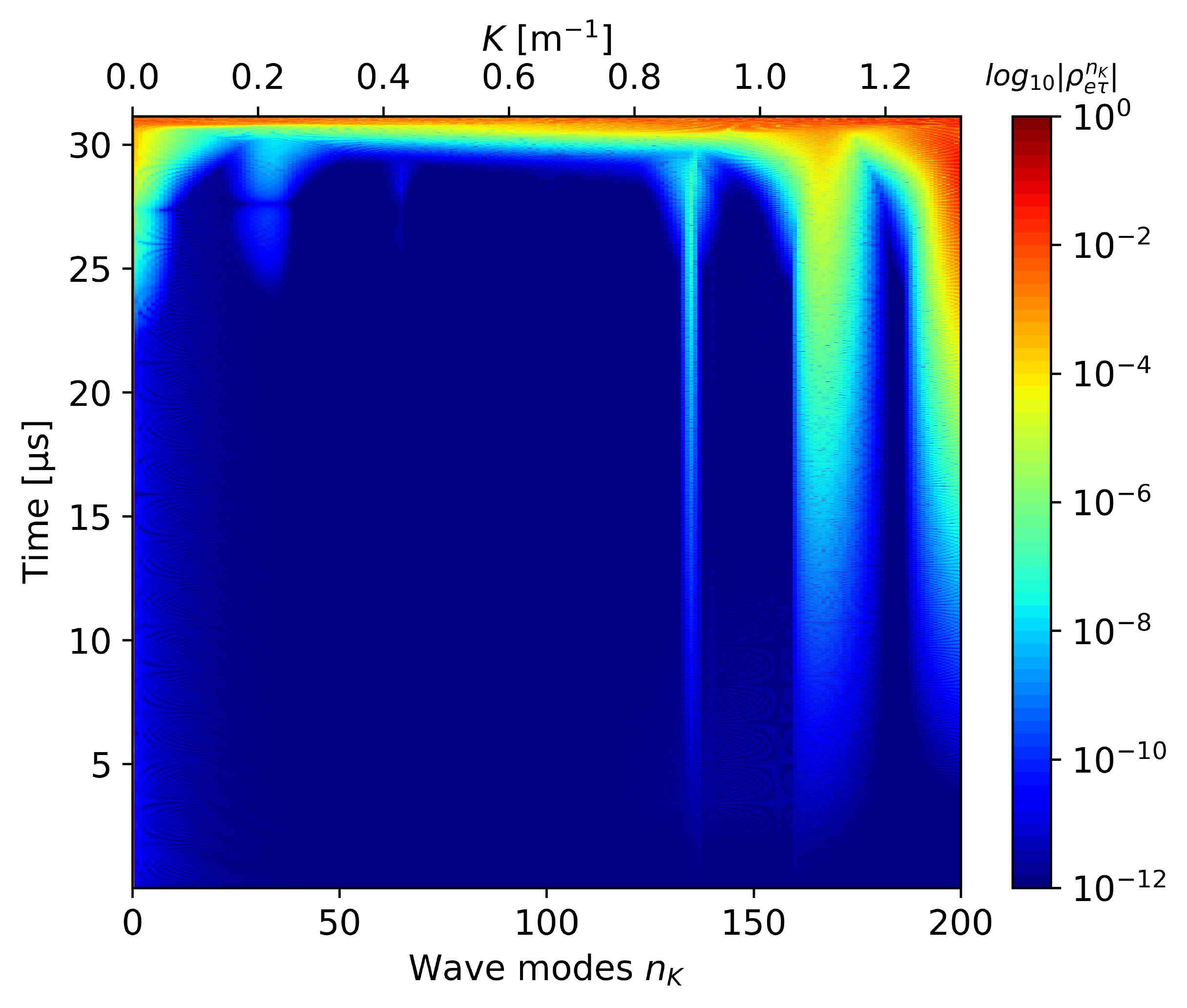}
    \includegraphics[width=0.32\linewidth]{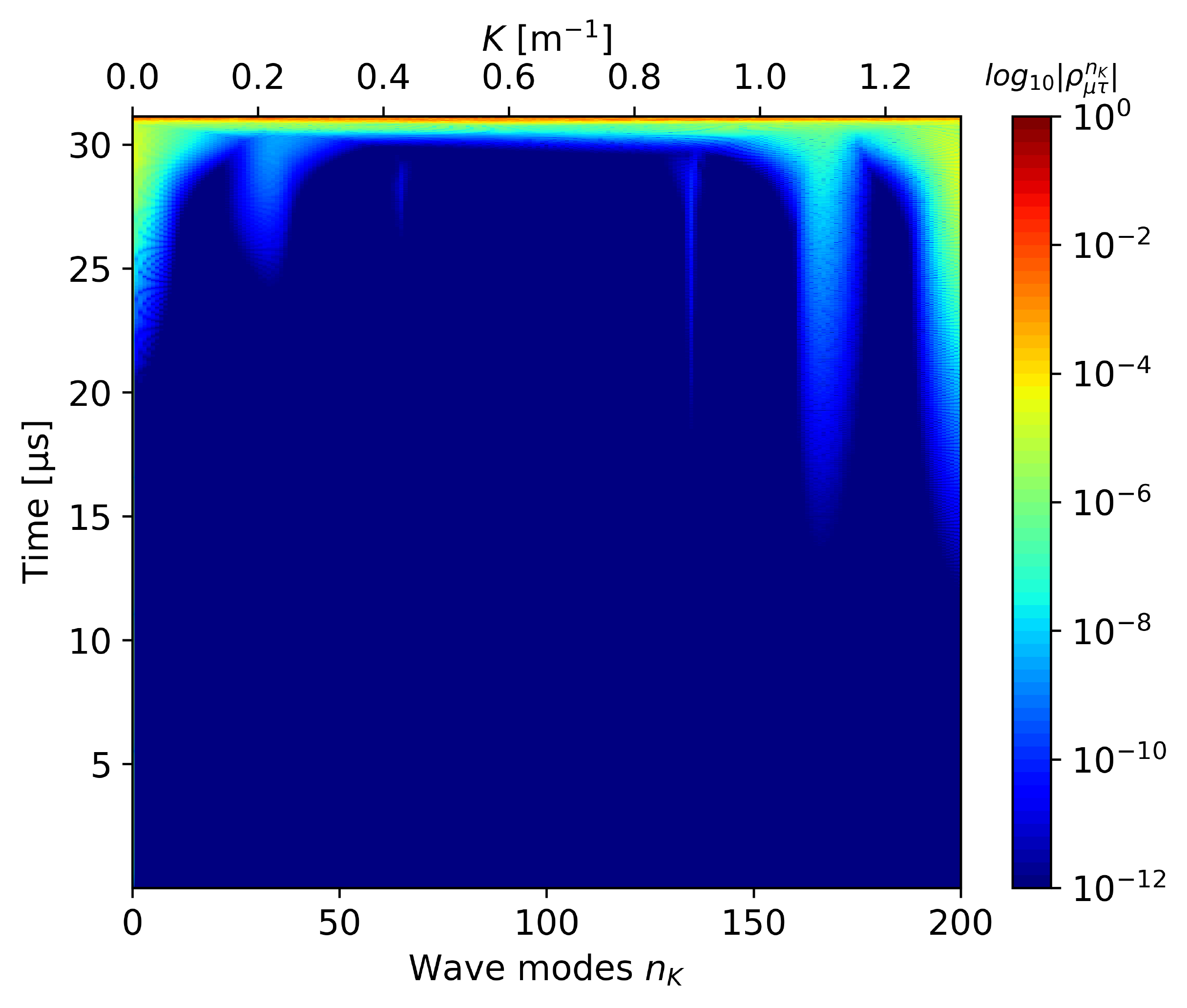}
    \caption{The same as Fig.~\ref{fig:3f_novac_nK}, but for the inclusion of the vacuum term.
    The flavor evolution appears to grow not only in the $e-\tau$ sector but in both the $e-\mu$ and $\mu-\tau$ sectors as well, which is different from the case ignoring the vacuum term in Fig.~\ref{fig:3f_novac_nK}.
    }
    \label{fig:3f_nK}
\end{figure*}
Next, we present the case within the three-flavor framework.
We perform the numerical simulation without the vacuum term to confirm the consistency with the linear stability analysis results.
Figure~\ref{fig:3f_novac_nK} shows the three off-diagonal components of the density matrix, $\rho^{e\mu}$, $\rho^{e\tau}$, and $\rho^{\mu\tau}$, which correspond to the NFLN differences, $G_{\boldsymbol{v}}^{e\mu}$,  $G_{\boldsymbol{v}}^{e\tau}$, and $G_{\boldsymbol{v}}^{\mu\tau}$.
In the three-flavor case, there is a crossing only in $G_{\boldsymbol{v}}^{e\tau}$ in our model, and it results in the exponential growth in the $e-\tau$ sector.
The nonlinear phase begins around $t=25\mathrm{~\mu s}$, which is much later than in the two-flavor case.
It is because the e-folding time predicted from the dispersion relation is about four times longer.
The unstable spatial modes in the linear phase of Fig.~\ref{fig:3f_novac_nK} match the dispersion relation in Fig.~\ref{fig:DR}.
After that, a cascade in spatial Fourier space starts to develop, and the flavor instability spreads into all spatial modes at $t\sim 35\mathrm{\mu s}$.

Similar to the two-flavor framework, we also take the vacuum term into account and perform the simulation.
The evolutions of all the off-diagonal terms are shown in Fig.~\ref{fig:3f_nK}.
The fast modes in the $e-\tau$ sector first grow as in the case neglecting the vacuum term.
The flavor instabilities in the $\mu-\tau$ and $e-\mu$ sectors are excited before a cascade starts to develop in Fourier space.
These behaviors are not found in the stability linear analysis, and the propagation of flavor instability across sectors is the effect of the vacuum mixing and the nonlinear term.
The fast modes in both the inert sectors can not be excited only by the corresponding NFLN difference.
The growth is caused by the flavor instabilities leaking out of the $e-\tau$ sector through the vacuum mixing.
Therefore, the same spatial modes $K$ evolve in the $e-\mu$ and $\mu-\tau$ sectors in Fig.~\ref{fig:3f_nK}.
On the other hand, a slight difference between the vacuum mixing in the sectors results in the small difference, as seen in Fig.~\ref{fig:3f_nK}.
Also, unexpected flavor instability emerges at spatial modes $K>1.2\mathrm{~m^{-1}}$, different from the case ignoring vacuum term.
The same as the effective two-flavor case in Sec.\ref{Sec.3A}, it may result from the existence of slow modes \cite{Airen:2018}.
In the $e-\tau$ sector, the origin $k=0$ is $K=\Phi^{e\tau}_r\sim 0.924\mathrm{~m^{-1}}$ and very weak flavor instability actually rises next to unstable modes corresponding to the middle branch.
Unstable modes at larger $K$ may also result from the mixing between fast and slow modes.

\begin{figure}[ht]
    \centering
    \includegraphics[width=0.9\linewidth]{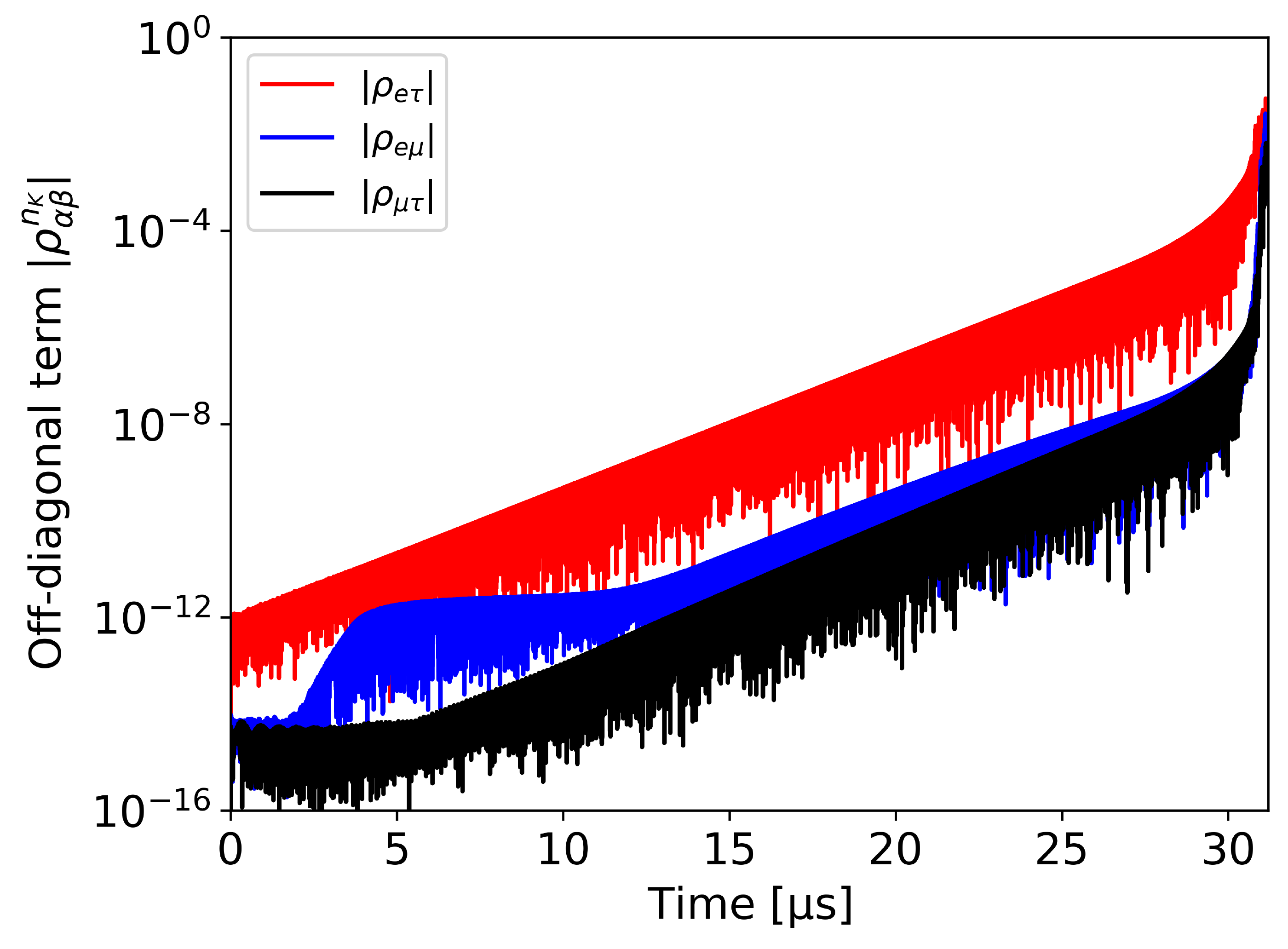}
    \caption{The same as Fig.~\ref{fig:novac_growth} but for the three-flavor framework including the vacuum term.
    Red, blue, and black lines are for the $e-\tau$, $e-\mu$, and $\mu-\tau$ sectors, respectively.
    The flavor instability in the $e-\tau$ sector first grows, and then those in $e-\mu$ and $\mu-\tau$ sectors evolve.
    }
    \label{fig:3f_growth}
\end{figure}
Figure~\ref{fig:3f_growth} shows the time evolution of the off-diagonal term $|\rho_{e\tau}|,|\rho_{e\mu}|,|\rho_{\mu\tau}|$ for $K=1.089\mathrm{~m^{-1}}$ in each sector in the case including the vacuum term.
The flavor instability in the $e-\tau$ sector first grows, and then those in the $\mu-\tau$ and $e-\mu$ sectors start to evolve.
The onset time of the linear growth in the inert sectors is later than the $e-\tau$ sector, while the growth rates are almost same.
This is due to the difference among the mass term in the three sectors.
The three-flavor effects via the vacuum mixing have been reported in Ref.~\cite{Shalgar:2021b}. 
The propagation of flavor instability across sectors is derived by flavor mixing in the vacuum term.
\\

\section{Conclusions}
\label{Sec4}
In this work, we have studied the nonlinear simulation on fast flavor conversion driven by tiny NFLN crossings in the preshock region of core-collapse supernovae.
We have exhibited that the unstable fast modes predicted in the linear regime are indeed excited by solving the equation of motion decomposed into spatial Fourier modes.
We have found that the convolution term in the nonlinear regime drives spatial modes that would otherwise remain stable and a cascade develops in Fourier space after perturbations grow sufficiently.
Besides, we have shown that the flavor instability starting up in the $e-\tau$ sector propagates across sectors within the three-flavor framework considering $\mu$LN and $\tau$LN angular distribution.

In the linear stability analysis, we have omitted the vacuum term and shown unstable modes derived from the fast instability.
However, the slow instability associated with the vacuum term may influence fast modes because the self-interaction term in the preshock region is not large enough to neglect the vacuum frequency completely.
The spatial Fourier modes have indeed demonstrated different flavor evolution compared to only the fast instability.
Our findings suggest that more generic studies for several situations are required to clarify the flavor instability in the presence of the vacuum term.

Recent works have pointed out that the consideration of heavy lepton flavor, especially muon neutrinos, greatly influences fast flavor conversion through the NFLN difference.
The angular distribution model which we used accounts for the effects of muon production and weak magnetism in the SN dynamics, and the enhanced muon antineutrino emission cancels out the ELN crossing.
On the other hand, the flavor instability in the $e-\tau$ sector may still survive under the environments and significantly influence the flavor conversion.
However, even this crossing may disappear if the weak-magnetism correction is more prominent than we assume.
It is crucial to clarify the three-flavor collective flavor conversion rooted in more realistic SN dynamics.
\\

\begin{acknowledgments}
We thank K. Sugiura for valuable discussions on determining the angular distribution of heavy lepton-flavor neutrinos.
M.Z. and T.M. are supported by the Japan Society for Promotion of Science (JSPS) Grant-in-Aid for JSPS Fellows (Grants No. 20J13631 and No. 19J21244) from the Ministry of Education, Culture, Sports, Science, and Technology (MEXT) in Japan.
Numerical computations were in part carried out on Cray XC50 at the Center for Computational Astrophysics, National Astronomical Observatory of Japan.
\end{acknowledgments}

\nocite{*}
%

\end{document}